\documentstyle[11pt]{article}
\input epsf
\def\sdtimes{\mathbin{\hbox{\hskip2pt\vrule
height 4.1pt depth -.3pt width .25pt\hskip-2pt$\times$}}}

\begin{document}
\thispagestyle{empty}
\baselineskip 20pt
\rightline{KIAS-P00051}
\rightline{KCL-TH-00-43}
\rightline{SNUTP-00-019}
\rightline{{\tt hep-th}/0008092}
\

\def\sdtimes{\mathbin{\hbox{\hskip2pt\vrule
height 4.1pt depth -.3pt width .25pt\hskip-2pt$\times$}}}

\def\tr{{\rm tr}\,}
\newcommand{\IR}{\relax{\rm I\kern-.08em R}}
\newcommand{\IZ}{\relax\ifmmode\mathchoice
{\hbox{\cmss Z\kern-.4em Z}}{\hbox{\cmss Z\kern-.4em Z}}
{\lower.9pt\hbox{\cmsss Z\kern-.4em Z}}
{\lower1.2pt\hbox{\cmsss Z\kern-.4em Z}}\else{\cmss Z\kern-.4em
Z}\fi}
\newcommand{\beq}{\begin{equation}}
\newcommand{\eeq}{\end{equation}}
\newcommand{\beqn}{\begin{eqnarray}}
\newcommand{\eeqn}{\end{eqnarray}}
\newcommand{\be}{\begin{eqnarray}}
\newcommand{\ee}{\end{eqnarray}}
\newcommand{\nn}{\nonumber}
\newcommand{\bde}{{\bf e}}
\newcommand{\ft}[2]{{\textstyle\frac{#1}{#2}}}
\newcommand{\eqn}[1]{(\ref{#1})}
\newcommand{\balpha}{{\mbox{\boldmath $\alpha$}}}
\newcommand{\bsalpha}{{\mbox{\small\boldmath$alpha$}}}
\newcommand{\bbeta}{{\mbox{\boldmath $\beta$}}}
\newcommand{\btau}{{\mbox{\boldmath $\tau$}}}
\newcommand{\blambda}{{\mbox{\boldmath $\lambda$}}}
\newcommand{\bepsilon}{{\mbox{\boldmath $\epsilon$}}}
\newcommand{\bphi}{{\mbox{\boldmath $\phi$}}}
\newcommand{\bnabla}{{\mbox{\boldmath $\nabla$}}}
\newcommand{\bpi}{{\mbox{\boldmath $\pi$}}}
\newcommand{\bX}{{\mbox{\boldmath $X$}}}
\newcommand{\ggg}{{\boldmath \gamma}}
\newcommand{\ddd}{{\boldmath \delta}}
\newcommand{\mmm}{{\boldmath \mu}}
\newcommand{\nnn}{{\boldmath \nu}}

\newcommand{\bra}[1]{\langle {#1}|}
\newcommand{\ket}[1]{|{#1}\rangle}
\newcommand{\sn}{{\rm sn}}
\newcommand{\cn}{{\rm cn}}
\newcommand{\dn}{{\rm dn}}
\newcommand{\diag}{{\rm diag}}


\centerline{\LARGE\bf The moduli space of two U(1) instantons }
\centerline{ \LARGE\bf on noncommutative $R^4$ and $R^3\times S^1$.}

\vskip 1.5cm
\centerline{\large\it 
 Kimyeong Lee$^*$\footnote{klee@kias.re.kr}, David 
Tong$^\dagger$\footnote{tong@mth.kcl.ac.uk}, and   Sangheon
Yi$^\ddag$\footnote{shyi@phya.snu.ac.kr}  } 
\vskip 1mm

\centerline{$^*$School of Physics, Korea Institute for Advanced Study}
\centerline{207-43 Cheongriangri-Dong, Dongdaemun-Gu}
\centerline{Seoul 130-012, Korea}

\centerline{$^\dagger$Department of Mathematics, Kings College, The Strand}
\centerline{London, WC2R 2LS, UK}

\centerline{$^\ddag$Department of Physics and Center for Theoretical Physics}
\centerline{Seoul National University, Seoul 151-742, Korea}
\vskip 3mm
\vskip 3mm

\vskip 1.cm
\begin{quote}
{\baselineskip 16pt We employ the ADHM method to derive the moduli
space of two instantons in $U(1)$ gauge theory on a noncommutative
space. We show by an explicit hyperK\"ahler quotient construction that
the relative metric of the moduli space of two instantons on $R^4$ is
the Eguchi-Hanson metric and find a unique threshold bound state.  For
two instantons on $R^3\times S^1$, otherwise known as calorons, we
give the asymptotic metric and conjecture a completion. We further
discuss the relationship of caloron moduli spaces of A, D and E groups
to the Coulomb branches of three dimensional gauge theory. In
particular, we show that the Coulomb branch of $SU(2)$ gauge group
with a single massive adjoint hypermultiplet coincides with the above
two caloron moduli space.}
\end{quote}

\newpage
\setcounter{footnote}{0}

\section{Introduction and Conclusion}

Recently there has been a great deal of interest in field theories on
non-commutative spaces and their classical soliton
solutions~\cite{conne,sw}.  Examples include instantons, monopoles,
vortices, CP(n) solitons, as well as novel solitons which only exist
in non-commutative
theories~\cite{nekra,hasi,gros0,india,yang,stro0}. While most recent
studies have focused on the characteristics of the solitonic field
configurations themselves, there have also been several works
investigating the low energy dynamics and, in particular, the moduli
spaces of these solitons. A key point is that the moduli spaces are
ordinary Riemann spaces; they feel the effect of the non-commutivity
of the underlying spacetime only through changes to their metric.

For example, the moduli space of a single instanton in a $U(n)$ gauge
theory on non-commutative $R^4$ or $R^3\times S^1$ has been identified
by one of us (K.L.), together with P.Yi in Ref.~\cite{pyi}.  For a
single instanton on $R^3\times S^1$ - otherwise known as a caloron -
simplifications occur when a Wilson line on the circle breaks the
gauge group to the maximal torus. In this case, the asymptotic form of
the moduli space may be derived by considering the dynamics of
constituent monopoles from which the caloron is constructed
\cite{caloron}. The resulting metric includes several parameters
corresponding to the radius, $R$, of the $S^1$ factor, the value of
the Wilson line, as well as the non-commutivity parameter itself.  For
all values of these parameters, the metric is hyperK\"ahler and
complete and, in analogy with monopoles, is believed to be
exact. Moreover, the metric has a smooth limit in the
decompactification limit $R\rightarrow\infty$. After separating the
center-of-mass motion, the remaining $4(N-1)$-dimensional metric is
known as the Calabi metric: as expected of the non-commutative $U(N)$
instanton moduli space, it has a tri-holomorphic $SU(N)$ isometry,
together with a non tri-holomorphic $U(1)$ isometry.  For the case of
$N=2$, the Calabi metric coincides with the Eguchi-Hanson metric,
which arises as the blow-up of $R^4/Z_2$.

In this paper, we further explore the moduli spaces of instantons.  We
firstly consider two instantons in a non-commutative $U(1)$ gauge
theory. In the ordinary, commutative limit, the instantons become
singular and their moduli space is simply the space of two unordered
points on $R^4$. After separating the center-of-mass motion, this
leaves the relative moduli space $R^4/Z_2$. The effect of the
non-commutative spacetime is to resolve this singularity and the
metric on the relative moduli space of two $U(1)$ instantons is again
expected to be Eguchi-Hanson\footnote{This point was also made
recently in \cite{ganor}.}.  In the first part of this paper,
we confirm this conclusion by performing an explicit hyperK\"ahler
quotient construction of the metric using the ADHM method. Note that
in this case, the quotient is performed with a non-abelian group.
We also find that there exists a unique threshold bound state of
these two U(1) instantons, in context of $N=4$ supersymmetric 
five dimensional theory. When six dimensional (1,1) supersymmetric
theory is compactified on a circle, a single instanton can be
regarded as a mode of winding number one. The threshold state
can be regarded as a mode of winding number two.

For two $U(1)$ non-commutative instantons on $R^3\times S^1$, we
derive the asymptotic form of the metric and observe that it coincides
with the $U(1)$ hyperK\"ahler quotient of the relative eight
dimensional moduli space for one massless and two massive monopoles in
$SO(5)$ gauge theory, whose metric was found explicitly sometime ago
by one of us (K.L) and C. Lu~\cite{lu}. In addition, both metrics 
become flat as the deformation parameter vanishes.  We conjecture that
this correspondence extends to the full metric.

We further describe how these metrics appear as the quantum corrected 
Coulomb branches of certain ${\cal N}=4$ three dimensional 
gauge theories. In particular, we show that the Coulomb branch of 
$SU(N)$ with a single massive adjoint 
hypermultiplet\footnote{After submitting this paper we were informed 
that this observation was previously made by Kapustin and 
Sethi \cite{kapseth}.} is given 
by the moduli space of $N$ $U(1)$ instantons on $R^3\times S^1$. 
The gauge coupling constant determines the radius $R$ of the 
$S^1$. Moreover, we describe the gauge theories yielding the 
moduli space of calorons for all simply-laced gauge groups and 
explain how these results tie in with known ideas of mirror 
symmetry in three dimensional gauge theories. 

The plan of this paper is as follows. In Sec.2, we briefly review the
ADHM formalism~\cite{adhm}. In Sec.3, the moduli space metric of two
$U(1)$ instantons on non-commutative $R^4$ is obtained explicitly by a
hyperK\"ahler quotient construction. In section 4 we discuss the
moduli space of two non-commutative calorons.  In section 5, we
discuss the vacuum moduli spaces of three dimensional gauge theories
and their relationship to the moduli spaces of various classical
solitons.

\section{The ADHM Formalism}

The ADHM approach to instanton physics arises naturally 
as the effective field theory describing $k$ D0 branes near $n$
parallel D4 branes in type IIA string theory. The low energy dynamics
of the D4 branes is described by a $U(n)$ gauge theory in 4+1 dimensions 
with 16 supercharges. The D0 branes appear as instantons within 
the D4 branes. The dynamics of D0 branes is given by quantum mechanics 
model with 8 supersymmetries which can be obtained from the dimensional 
reduction of $N=1$ supersymmetric 6 dimensional gauge theory with gauge 
group $U(k)$, one adjoint hypermultiplet and $n$ fundamental
hypermultiplets. The D-terms describing the Higgs branch of the 
D0-brane theory coincide with the ADHM constraints and, after 
modding out by the $U(k)$ gauge symmetry, give a hyperK\"ahler 
quotient description of the instanton moduli space. We denote the 
moduli space of $k$ $U(n)$ instantons as ${\cal M}_{k,U(n)}$. 

When a uniform external NS-NS $B_{\mu\nu}$ field is present, with 
non-vanishing components of both indices along the longitudinal directions 
of the D4 branes, the field theoretic limit is the five dimensional gauge
theory with the same gauge group and supersymmetry, but with the 
spatial $R^4$ becoming non-commutative,
\beq
[x^\mu, x^\nu] = i \theta^{\mu\nu}\; ,
\eeq
where $\mu,\nu=1,\cdots,4$ and $B_{\mu\nu}^{-1}\sim\theta_{\mu\nu}$.  When
$B_{\mu\nu}$ has a self-dual part, the D0 branes (which appear as
anti-self-dual instantons) cannot escape the D4-branes, a fact which is
reflected in the disappearance of the singularities in the instanton
moduli space. {}From the perspective of the D0-brane gauge theory, the
self-dual part of $B_{\mu\nu}$ induces a FI parameter.  This modifies
the classical Higgs branch metric, blowing up its singularities.  We
consider the selfdual case where nonvanshing components of
$\theta^{\mu\nu}$ are
\beq
\theta^{12} = \theta^{34} = -\frac{\zeta}{4}\; .
\eeq
Defining,
\beq
y^0 = x^4+ix^3, \;\; y^1 =-i( x^1-i x^2)\; ,
\eeq
we have the non-vanishing commutation relations,
\beq
[y^0, \bar{y}^0] =  [ y^1,\bar{y}^1] = \frac{\zeta}{2}\; .
\eeq
The instanton moduli space, whether deformed by FI term or not, is a
hyperK\"ahler space and the metric has three complex structures and
corresponding Kahler forms. The ADHM construction provides the standard
method to find the metric and Kahler forms, both of which arise 
from the kinetic terms for the matter hypermultiplets in the $U(k)$ 
theory of the D0 brane world-volume theory. To find the  moduli 
space of $k$ instantons in $U(n)$ gauge theory, we
start from two $k$ dimensional complex square matrices $B_0,B_1$ and
two $n$ by $k$ complex matrices $I^\dagger$ and $J$. They form $ 4k^2
+4kn$ dimensional flat hyperK\"ahler space ${\cal M}_0$ with the metric
\beq
ds^2 = 2\; {\rm tr}_k( dB_0 \otimes_s dB_0^\dagger + dB_1 \otimes_s
dB^\dagger_1 + dI\otimes dI^\dagger + dJ^\dagger\otimes dJ) \; ,
\label{metric1}
\eeq
where $\otimes_s$ is the symmetic direct product. In this space, there
exist three constant complex structures ${\cal I}_{s}$ with $s=1,2,3$
such that ${\cal I}_r {\cal I}_s = -\delta_{rs} 1_{4k^2+4kn} +
\epsilon_{rst} {\cal I}_r$, and the corresponding K\"ahler forms,
\beqn
&& w_3= i\; \tr_k( dB_0\wedge dB_0^\dagger +dB_1\wedge dB_1^\dagger + dI
\wedge dI^\dagger - dJ^\dagger \wedge dJ ), \nonumber\\
&& w_1 - i w_2 = 2i \; \tr_k( dB_0\wedge dB_1 + dI \wedge dJ) \; .
\label{kahler1}\eeqn
(For the given metric $ds^2=g_{ab}dx^a dx^b$ and the covariantly
constant complex structure $I^{\; a}_{(s)b}$, one constructs the Kahler form
$ w_{s}= \frac{1}{2} I_{(s)ab} dx^a\wedge dx^b$.)

There is a gauge group $U(k)$ on this space, under which $B_0,B_1$
belong to the adjoint representation and $I, J^\dagger$ to the
fundamental representation.  There are also three natural moment maps
\beqn
&& \mu_r = [B_0,B_0^\dagger]+[B_1,B_1^\dagger]+ I I^\dagger -
J^\dagger J , \nonumber \\ 
&& \mu_c = [B_0,B_1] + IJ \; ,
\eeqn
which map from ${\cal M}_0$ to the coadjoint orbit of $U(k)$.  The
moduli space of instantons on commutative space is given by the
hyperK\"ahler quotient, ${\cal M}_1 = \{ \mu_r^{-1}(0)\cap
\mu_c^{-1}(0) \}/U(k)$. This moduli space of dimension $4kn$ is
singular at some points where some of instantons shrink to zero
size. The moduli space of instantons on non-commutative space is a
deformation of this moduli space due to the parameter $\zeta>0$ and is
defined by the hyperK\"ahler quotient~\cite{nekra,naka}
\beq
{\cal M}_{k,U(n)} = \{\mu_r^{-1}(\zeta 1_k)\cap \mu_c^{-1}(0) \} /
U(k)
\; .
\eeq
This space is no longer singular, reflecting the fact that instantons 
cannot shrink to zero 
size. The metric and complex structures on the instanton moduli space are
naturally induced from the original metric (\ref{metric1}) and complex
structures (\ref{kahler1}) by  the hyperK\"ahler quotient process.
This is the structure we are interested in.

For a given set of matrices $B_0,B_1, I, J$, one can 
define a  (n+2k) by 2k matrix
\beq
\Delta(x) = \left( \begin{array}{cc}
                    I^\dagger & J \\ 
             B_0^\dagger -\bar{y}_0 & B_1 -y_1 \\
	     -B_1^\dagger + \bar{y}_1 & B_0-y_0 
  		\end{array} \right) \; .
\eeq
The equations  $\mu_r=\zeta, \mu_c=0$ are identical to the ADHM constraint
\beq
\Delta^\dagger \Delta = f^{-1}(x)\otimes 1_2 \; ,
\eeq
where $f$ is a nonsingular k dimensional square matrix.
For each solution to the ADHM constraint, 
one can find  $n+2k$ by $n$ matrix $v$ satisfying the equation
\beq
\Delta^\dagger v(x) = 0 \; ,
\eeq
together with the normalization condition
\beq
v^\dagger v = 1_n \; .
\eeq
The selfdual gauge field configuration is then
\beq
A = v^\dagger dv \; .
\eeq
The gauge field and field strength appear as operators defined on the
Hilbert space specified by creation operators and annihilation
operators made of spatial coordinate. Even though their moduli space
is a hyperK\"ahler space without any singularity in the sense of the
Riemann geometry, the gauge field strength, which is an operator on
the Hilbert space, is still singular and needs some sort of
projection, or a modification of the underlying background space. We
feel this issue is not settled and needs further study~\cite{branden}.

\section{Two Instantons on Non-Commutative $R^4$}

For a single instanton of the $U(N)$ gauge theory, the moduli space on
$R^4$ is made of the center of mass $R^4$, together with the $4(N-1)$
dimensional Calabi metric which describes the scaling and gauge
symmetry~\cite{pyi}. The next most simple example is that of two
identical $U(1)$ instantons on non-commutative $R^4$. To find their
moduli space metric, we start with finding the ADHM data satisfying
the ADHM constraint,
\beq
\mu_r = \zeta 1_k \; , \;\;\; \mu_c=0\; ,
\label{constr}
\eeq
for two instantons in U(1) gauge theory.  The general solution for
this equation with two instantons is known in another 
context~\cite{furuu1}.  The step behind this is rather simple. First, one
takes the trace of these equation to get constraints on $I,J$, which
are $J \cdot I=0$ and $I^\dagger I - JJ^\dagger = 2\zeta$. One may 
use the $U(2)$ gauge symmetry to set $I^\dagger = (\alpha,0) $ and 
$J= (0,\beta)$. These two equations show that the only consistent solution 
has $J=0$. This now restores the $U(2)$ symmetry and it may be employed 
once more, this time to simultaneously triangularize $B_0,B_1$. 
Solving  the equations with triangularized $B_0,B_1$ and
$I$ yields the ADHM data with satisfies \eqn{constr}, given 
by $J=0$ and
\beqn
&& B_0 = w_0 1_2 +\frac{z_0}{2} \left( \begin{array}{cc}
               1 & \sqrt{\frac{2b}{a}}  \\ 0 & -1
\end{array} \right) \; , \nonumber  \\ 
&& B_1 = w_1 1_2+ \frac{z_1}{2} \left( \begin{array}{cc} 
               1 & \sqrt{\frac{2b}{a}}  \\ 0 & -1
               \end{array} \right) \; , \nonumber  \\ 
&& I = \sqrt{\zeta} \left( \begin{array}{c}
                \sqrt{1-b} \\ \sqrt{1+b} 
                \end{array} \right) \; , 
\eeqn
where we have chosen $I$ to be real using the remaining gauge 
transformation, and where the dimensionless parameter is given by  
\beqn
&& a= \frac{ |z_0|^2+ |z_1|^2}{2\zeta} \ge 0 \; , \nonumber  \\
&& b = \frac{1}{a+\sqrt{1+a^2}} \le 1  \; .
\eeqn
The parameters $w_0, w_1$ represent the center of the mass coordinate
of two instantons, and the parameters $z_0,z_1$ represent the relative
position between two instantons. They are the coordinates of the eight
dimensional moduli space for two instantons.

There are couple of comments to be made before we proceed to calculate
the metric on the moduli space. First, in the limit of large
separation, or large $a$, the matrices $B_0$ and $B_1$ become
diagonal. Their eigenvalues $w_i \pm z_i/2$ are the positions of the
two instantons. Second, we have not exhausted the gauge degrees of
freedom by choosing the above solution. Under an SU(2) transformation
\beq
U= i\left( \begin{array}{cc}
	 -b & \sqrt{2ab} \\
	\sqrt{2ab} & b 
	\end{array} \right) \; , 
\eeq
the relative coordinates $z_0, z_1$ change their sign and the matrix
$I$ is invariant. Thus the exchange of the two instanton positions is
a part of the gauge transformation, by which we want to mod out. This
is expected as two instantons are identical solitons and so their
relative moduli space should be $R^4/Z_2$ in large separation. Third,
in the limit $a=0$ where the two instantons overlap, the relative
parts of $B_0$ and $B_1$ have non-vanishing upper right corner, and
$I$ has a nonzero lower component. Thus, under $U(2)$ transformation
\beq
U= \left( \begin{array}{cc} 
		 e^{i\beta} & 0 \\
		0 & 1  \end{array}\right) \; ,
\eeq
the relative data space becomes $ (z_0/\sqrt{a}, z_1/\sqrt{a})\sim
e^{i\beta} (z_0/\sqrt{a}, z_1/\sqrt{a}) $. This $U(1)$ gauge group
makes the data space a two space, $S^3/U(1) = S^2$ .

The tangent vectors for this  ADHM data are  a sum of the
differentials with eight parameters and infinitesimal gauge
transformation, 
\beqn
&&\delta B_0 = dB_0 - i[\alpha,B_0] \; , \nonumber  \\
&& \delta B_1= dB_1 -i[\alpha,B_1] \; \nonumber \\
&& \delta I = dI -i \alpha I \; , \nonumber \\
&& \delta J^\dagger = dJ^\dagger -i\alpha J^\dagger \; , 
\eeqn
where $\alpha$ is a hermitian two by two matrix and  the differential 
$d$ acts on  all eight parameters. The tangent vectors satisfy the
linearized ADHM constraints automatically. In addition, they are also 
required to satisfy the linearized Gauss law,
\beq
[\delta B_0, B_0^\dagger] - [B_0, \delta B_0^\dagger] +
[\delta B_1, B_1^\dagger] - [B_1, \delta B_1^\dagger] +
\delta I I^\dagger - I \delta I^\dagger -
\delta J^\dagger J + J^\dagger \delta J = 0  \; .
\eeq
These  equations  can be put 
together into
\beq
[\delta B_0, B_0^\dagger] + [\delta B_1, B_1^\dagger] + \delta I
I^\dagger = 0 \; , 
\eeq
which fixes the matrix $\alpha$ uniquely,
\beq
\alpha = \frac{i b}{2\sqrt{1+a^2}} \left( \begin{array}{cc}
              \frac{\bar{\partial} a - \partial a}{1-b} & 
                   - \frac{\sqrt{2}\bar{\partial}a }{\sqrt{ba}} \\ 
                \frac{\sqrt{2}\partial a }{\sqrt{ba}} &
                   \frac{\bar{\partial}a -\partial a}{1+b} 
             \end{array}\right) \nonumber \; , 
\eeq
where $\partial a = (\partial a /\partial z_i)dz_i$ and
$\bar{\partial} a = (\partial a /\partial \bar{z}_i)d\bar{z}_i$.

Now it is straightforward to write the explicit expression for the
tangent vector. It turns out $\delta I = 0$ in our case and so the
metric of the moduli space for two instantons is,
\beq
ds^2 = 2 \tr( \delta B_0 \delta B_0^\dagger + \delta B_1 \delta
B_1^\dagger )  \; .
\eeq
The  metric naturally splits into to the center of
mass part and the relative motion part: 
\beq
ds^2=  ds^2_{\rm cm} + ds^2_{\rm rel} \; ,
\eeq
where
\beq
ds_{\rm cm}^2 = 2 dw_id\bar{w}_i \; ,
\eeq
and
\beq ds_{\rm rel} = \frac{\sqrt{1+a^2}}{a} dz_i d\bar{z}_i -
\frac{1}{ a^2\sqrt{1+a^2}} \frac{1}{2\zeta}\bar{z}_i dz_i z_j
d\bar{z}_j \; .
\eeq
Under the change of variables to radial and  angular 
coordinates $r, \theta,\varphi$ and $\psi$, 
\beqn
z_0= r \cos\frac{\theta}{2} \exp \frac{i}{2}(\psi +\varphi) \; , \nonumber \\
z_1= r \sin\frac{\theta}{2} \exp \frac{i}{2}(\psi -\varphi) \; ,
\eeqn
the relative metric becomes
\beq
 ds^2_{\rm rel} = \frac{1}{\sqrt{1 + \frac{4\zeta^2}{r^4} }}(dr^2 +
\frac{1}{4} r^2 \sigma_z^2) +  \frac{1}{4}\sqrt{1+
\frac{4\zeta^2}{r^4}} r^2(\sigma_x^2+ \sigma_y^2)  \; ,
\eeq
where $\sigma_i$ are the  standard left SU(2) invariant one forms 
\beqn
&& \sigma_x = -\sin \psi d\theta +\cos\psi \sin \theta d\varphi \; ,
 \nonumber \\
&& \sigma_y = \cos \psi d\theta +\sin \psi \sin \theta d\varphi \; ,
 \nonumber \\
&& \sigma_z = d\psi +\cos\theta d \varphi  \; , 
\eeqn
which satisfy $d\sigma_z = \sigma_x \wedge \sigma_y$ and cyclic permutations 
thereof. For the angular variables on $S^3/Z_2$, the ranges are given by, 
\beq
0\le \theta\le \pi, \; 0\le \varphi \le 2\pi, \; 0\le \psi \le 2\pi\ ,
\eeq
in contrast to the $S^3$ case for which $0\le \psi \le 4\pi$.

Near the origin $r\approx 0$, the relative metric in new radial
variable $\rho=r^2$ becomes
\beq
ds_{\rm rel} \approx \frac{1}{8\zeta} (d\rho^2 + \rho^2 d\psi^2) +
\frac{\zeta}{2} (d\theta^2 + \sin^2 \theta d\varphi^2) \; ,
\eeq
which is the metric for $R^2\times S^2$. As the range of $\psi$ is
$[0,2\pi]$, this metric is nonsingular.  Thus the metric is smooth in
the limit where two instantons are coming together.

As the relative space is hyperK\"ahler, there exist three complex structures
and corresponding Kahler forms. These three  K\"ahler forms becomes
\beqn
&& w_3 =\frac{1}{2} \left( \frac{ a}{\sqrt{1+a^2}}  rdr \wedge
\sigma_z + \frac{\sqrt{1+a^2}}{2a} r^2 \sigma_x\wedge \sigma_y \right)
\; , \nonumber \\
&& w_1-iw_2 = \frac{r}{2}(dr+ ir\frac{\sigma_z}{2})\wedge
(\sigma_x+i\sigma_y) 
\; .
\eeqn

The $SU(2)$ transformations which leave one-forms
$\sigma_x,\sigma_y,\sigma_z$ invariant, leave the metric and the
K\"ahler forms invariant and are thus a triholomorphic isometry. In
addition the $U(1)$ rotation, $\psi\rightarrow \psi + \epsilon$ leaves the
metric and the Kahler form $w_3$ invariant. However, it does not leave
$w_1 - iw_2$ invariant and so its action is not a triholomorphic
isometry. This $U(1)$ symmetry is the surviving symmetry of the SO(3)
symmetry which rotates the three FI parameters $\vec{\zeta}$.

This symmetry consideration suggests that one could have guessed the
metric of two U(1) instantons on non-commutative space should be
Eguchi-Hanson~\cite{eguchi}: the moduli space should be hyperK\"ahler
and the relative part should be four dimensional. Moreover,
asymptotically the relative metric should be $R^4/Z_2$. The rotational
symmetry O(4) of $R^4$ should be broken to $SO(3) \times U(1)$ due to
the FI term $\vec{\zeta}$.  In addition the $SO(3)$ symmetry should be
triholomorphic and the $U(1)$ homomorphic. There is only one such
hyperK\"ahler space, which is of course the Eguchi-Hanson metric.

With the change of coordinates,
\beq
u^4 = r^4 +4\zeta^2 \; , 
\eeq
we get the standard form of the Eguchi-Hanson metric
\beq ds^2 = \frac{du^2}{1- \frac{4\zeta^2}{u^4}} + \frac{u^2}{4}
\left( \sigma_x^2 + \sigma_y^2 + (1- \frac{4\zeta^2}{u^4}) \sigma_z^2
\right) \; .
\eeq 

In the five dimensional ${\cal N}=2$ Yang-Mills theory, the low-energy
dynamics of instantons inherits $N=8$ supercharges.  The ground state
of the supersymmetric quantum mechanics is then given by a
normalizable, self-dual harmonic form on the moduli
space~\cite{witten}. On the Eguchi-Hanson metric, there exists a
unique form with these properties,
\beqn
\Omega &=&  d\left( \frac{\sigma_z}{u^2} \right) \nonumber \\ 
 &=& \frac{1}{u^3} (-2du\wedge \sigma_z + u \sigma_x\wedge \sigma_y) 
\eeqn
modulo a constant factor. Moreover, this form is invariant under
$SO(3) \times U(1)$; it can be interpreted as a threshold bound state
of two $U(1)$ instantons on noncommutative space.  The existence of
this bound state has been advocated before in Ref.~\cite{rozali}.  The
instantons of the 16 supersymmetric five dimensional Yang-Mills theory
are identified with the Kaluza-Klein momentum modes of the six
dimensional (2,0) supersymmetric theory compactified on a circle.
Thus the threshold bound state of two $U(1)$ instantons corresponds to
the state of two KK momenta. (In the similar vein, this threshold
bound state can be also  interpreted as the single particle state for a
single $U(2)$ instanton on the noncommutative space, corresponding to
the state of the KK mode of the $(2,0)$ theory with $U(2)$ gauge
group.)

\section{Two Instantons on non-commutative $R^3\times S^1$}

In general, instantons, or calorons, on $R^3\times S^1$ with
nontrivial Wilson loop along the circle can be interpreted as a charge
neutral composite of constituent BPS monopoles \cite{caloron}.  This 
construction for general simply-laced gauge group 
will be reviewed in the following section.  While this interpretation 
in terms of monopoles is true for non-commutative 
$U(n)$ gauge theory with $n\ge2$, 
there is a subtlety in the $U(1)$ gauge theory \cite{pyi}: rather 
than a magnetic monopole, the caloron now appears as a magnetic 
dipole. This is most simply seen in the T-dual version of the 
D0-D4 system, in which calorons appear as D-strings wrapping the 
dual circle, ending on D3-branes. In the presence of a background 
NS-NS $B$-field, the BPS D-string is no longer perpendicular to the D3-branes, 
but is tilted \cite{hasi}. The string therefore spirals around the circle and 
its two ends lie at different points on the D3-brane, where it appears 
as a dipole.

More specifically, let us compactify the $x^4$ direction with radius $R$; 
$0\le x^4 \le 2\pi R$. The non-commutative ADHM equations 
then reduce to the non-commutative Nahm equation,
\beq
\frac{dT^i}{dt} - i [T_4, Ti] = \frac{i}{2} \epsilon_{ijk} [T^j,T^k] + 
\delta_{i3} \frac{\zeta}{2} + {\rm tr}_2 ( \tau_i a
a^\dagger)\delta(t-v)
\eeq
with $t$ in the interval $[-1/(2R),1/(2R)]$.  For a single caloron,
the function $T_i(t)$ is the position of the D string, and $T_3(t)$
increases linearly in $t$ with slope $\zeta/2$.  The end points where
this D string meets D3 branes appear as the positive and negative
monopoles, now with a finite separation of ${\zeta}/{2R}$
along the $x^3$ direction. Thus, a single caloron appears as a magnetic
dipole.  Now let us consider the dynamics of two dipoles: we denote
the relative position between two positive charge as ${\bf r}$. This,
of course, is identical to that between two negative charges. The
relative positions between positive and negative charges are then
${\bf r} \pm (\zeta/2R) \hat{\bf z}$. While it is not obvious how to
get the asymptotic form of the metric for large separation of monopoles
in the gauge theory on non-commutative space, let us calculate the
asymptotic metric in most naive way.  The asymptotic form of the
metric is then determined just by their charge.  {}From the T-dualized
D3-D1 brane picture, the identical charges have magnetic repulsion
and Higgs attraction, while the opposite 
charges with D1 strings coming out
from D3 branes oppositely, have magnetic attraction and Higgs
repulsion.

Thus, the metric in large separation obtained by the standard
technique~\cite{gibbons} becomes
\beq
ds^2 = R\left( U({\bf r}) d{\bf r}^2 + U^{-1}({\bf r}) (d\psi + {\bf W}({\bf
r}) \cdot d{\bf r})^2 \right) \; ,
\label{asympt}
\eeq
where 
\beq
U({\bf r}) =\frac{1}{R} - \frac{2}{|{\bf r}|} + \frac{1}{|{\bf r} + (\zeta/2R)
\hat{\bf z} |} +\frac{1}{|{\bf r} - (\zeta/2R)
\hat{\bf z} |} 
\eeq
up to overall constant.

A metric whose asymptotic form is identical to that of 
Eq.(\ref{asympt}) has appeared previously in the discussion of the
relative moduli space of one massless and two massive monopoles in
SO(5) gauge theory, which is spontaneously broken to $U(1) \times
SO(3)$ \cite{lu}.  The net magnetic charge is purely abelian and the
relative moduli space is eight dimensional. One can take the
hyperK\"ahler quotient of this eight dimensional metric with the
unbroken $U(1)$ triholomorphic gauge symmetry of unbroken $SO(3)$ to 
get a four dimensional metric. Physically one may think of this as
holding the position of the massless monopole fixed relative to the
center of mass. This four dimensional metric, found by one of us 
(K.L) and Lu, has the same asymptotic form as Eq.  \eqn{asympt}. 
We therefore conjecture that the full metric coincides with the relative 
moduli space of two U(1) calorons on non-commutative $R^3\times S^1$.  

There are several pieces of evidence for this conjecture.  Firstly,
notice that both metrics become flat in the zero deformation parameter
limit, which at least shows that our conjecture is consistent.
Secondly, as shown in Ref.~\cite{lu}, the monopole moduli space of
$SO(5)$ theory originates from that of two massive and two massless
monopoles in $SU(4)$ gauge theory, whose symmetry is broken to
$SU(2)\times U(1)\times SU(2)$. The identification of two massless
monopoles in the $SU(4)$ case leads to the $SO(5)$ case. Thus one can
get a two parameter family of four dimensional Hyperk\"ahler spaces by
independently holding fixed the two massless monopoles of the $SU(4)$
theory \cite{houghton}. For example, when one of the parameters is
sent to infinity, one naturally gets Dancer's metric \cite{dancer},
which describes two massive and one massless monopoles in $SU(3)
\rightarrow SU(2)\times U(1)$. When the remaining parameter is sent to
zero, this reduces to the double cover of the Atiyah-Hitchin. In
contrast, when the two parameters coincide and are nonzero, one gets
the four dimensional Lee-Lu metric that is of interest here.

The above discussion has a parallel in terms of three dimensional gauge
theories. As Seiberg and Witten~\cite{threemon} have observed, the 
$N=4$ supersymmetric three dimensional gauge theory with $SU(2)$ gauge group 
and one massive fundamental hypermultiplet has a quantum corrected 
Coulomb branch given by the Dancer metric. The case of interest here 
however is that of two, identified, fundamental hypermultiplets or, 
equivalently, a single hypermultiplet in the symmetric 
representation. For $SU(2)$, this is equivalent to a single massive 
adjoint hypermultiplet. As will be discussed in the following section, 
the Coulomb branch of this theory indeed coincides with the 
moduli space of two $U(1)$ non-commutative  calorons. Related 
studies ALF spaces with the blow up of ADE singularities with the asymptotic
space $R^3\times S^1/\Gamma$, where $\Gamma$ is the subgroup of
$SU(2)$ can be found in~\cite{kap}.

One of  the interesting questions is how many threshold bound states
two U(1) calorons have. In the $R^4$ limit, there exists one as
discussed in the previous section. It remains to be understood 
whether there are additional bound states which disappear in $R^4$ limit.

\section{Calorons, Instantons and Three Dimensional Gauge Theories}

We now turn to the construction of the moduli spaces of calorons and 
instantons as the quantum corrected Coulomb branch of three 
dimensional gauge theories. We will derive the gauge theories whose 
Coulomb branches yield the moduli space of calorons in $A$,$D$ and 
$E$ gauge groups. We will further show that in the case of $A$ gauge 
groups only, there exists a mass parameter which deforms the 
Coulomb branch into the moduli space of non-commutative calorons. 
The limit in which the caloron moduli space becomes the instanton 
moduli space is shown to be the strong coupling limit of the 
gauge theory. 

Let us start by recalling a few facts 
about instantons on ordinary (commutative) $R^3\times S^1$. 
Specifically, we restrict attention to calorons in a 
simply-laced gauge group $H$ which is broken to the maximal torus by a 
Wilson line around $S^1$. It was shown in \cite{caloron,kimyeong} that 
the resulting configurations are constructed of fundamental magnetic 
monopoles associated to the nodes of the {\em extended} Dynkin diagram of $H$. 
To each of the nodes of the Dynkin diagram, there is an associated 
root $\vec{\beta}_i$, $i=0,\cdots,r=\,{\rm rank}\,(H)$. For $i=1,\cdots,r$, 
these are the simple roots of $H$, while the lowest root of $H$, 
$\vec{\beta}_0$, is associated with 
the extended node. 
The topological charges of the caloron, which include both magnetic charge 
and Pontyagrin number, are determined by the choice 
of the number of monopoles of each type: these numbers are denoted 
by $n_i>0$, $i=0,\cdots,r$. In 
particular, the configurations which become the $k$-instanton solution 
in the decompactified $R^4$ limit are given by $n_i=k d_i$, where 
$d_i$ are the Dynkin indices of $H$ \cite{kimyeong}\footnote{Recall 
that for $A_r$, $d_i=1$ for all $i$, for $D_r$, $d_i=1$ for $i=0,1,r-1,r$ 
and $d_i=2$ for $i=2,\cdots,r-2$. Finally, for $E_6$, $d_i=1,1,1,2,2,2,3$, 
for $E_7$, $d_i=1,1,2,2,2,3,3,4$ and for $E_8$, $d_i=1,2,2,3,3,4,4,5,6$.}.

The three dimensional gauge theory which will reproduce the moduli 
space of calorons has ${\cal N}=4$ 
supersymmetry (8 supercharges). The gauge group ${\cal G}$ and 
and the hypermultiplet matter content is determined by 
a quiver diagram based on the extended Dynkin diagram of $H$,
\beq
{\cal G}=\prod_{i=0}^rU(n_i) \; .
\nonumber 
\eeq
The gauge coupling constant of the $U(n_i)$ factor is $e_i$.  Whenever
there exists a line joining the $i^{\rm th}$ and $j^{\rm th}$ node
(i.e. whenever $\vec{\beta}_i\cdot\vec{\beta}_j\neq 0$ for $i\neq j$),
the three-dimensional gauge theory also includes a hypermultiplet
transforming in the $({\bf n}_i,\bar{{\bf n}}_j)$ bi-fundamental
representation of $U(n_i)\times U(n_j)\subset{\cal G}$.

For the case in which $n_i=d_i$, where $d_i$ are the Dynkin indices of $H$, 
the gauge theory above coincides with the Kronheimer gauge group which 
has been discussed in the context of mirror symmetry in three 
dimensional gauge theories in \cite{is}. For the case in 
which $n_i=kd_i$ for $k>1$, the gauge theory is very similar 
to those discussed in \cite{berk}. We will comment on this at 
the end of this section. Further discussion on the relationship between 
caloron moduli spaces and the vacuum moduli spaces of gauge theories 
can be found in \cite{sethi,kapseth}.

We are interested in computing the low-energy dynamics on the Coulomb
branch of the theory. In such vacua the vacuum expectation values
(VEVs) of the hypermultiplet scalars are set to zero, while the
triplet of scalars, ${\bphi}$, that live in the vector multiplet
acquire a VEV which, generically, breaks the gauge group ${\cal G}$ to
its maximal torus. We assume such symmetry breaking does indeed
occur. The massless fields on the Coulomb branch thus consist of $3N$
real scalars, $\bphi^a$, $a=1,\cdots,N$ and $N$ photons, where
$N=\sum_{i=0}^rn_i$. Dualising the photons into $N$ periodic scalars,
$\sigma^a$, which we take to have period $2\pi$, we are left with a 
sigma-model with a $4N$ dimensional
target space. Supersymmetry requires the metric on this space to be
hyperK\"ahler. Although classically flat, this metric receives
one-loop corrections, as well as non-perturbative corrections due to
various instanton effects. While the latter are generally difficult to
calculate, the former result in a toric hyperK\"ahler metric given by,
\beq
ds^2 = g_{ab}d\bphi^a\cdot d\bphi^b+(g^{-1})^{ab}
\left(d\sigma_a+{\bf W}_{ac}\cdot d\bphi^c\right) \left(d\sigma_b+{\bf
W}_{bd}\cdot d\bphi^d\right) \; 
\label{coul}
\eeq
with
\beqn
g_{aa} &=& \frac{1}{\tilde{e}_a^2}-\sum_{b\neq
a}\frac{\vec{\alpha}_a\cdot\vec{\alpha}_b} 
{|\bphi^a-\bphi^b|} \; , \nonumber\\
g_{ab}&=&\frac{\vec{\alpha}_a\cdot\vec{\alpha}_b}{|\bphi^a-\bphi^b|}\ \ \ \  
(a\neq b) \; , \nonumber
\eeqn
where $\vec{\alpha}_a=\vec{\beta}_i$ and $\tilde{e}_a=e_i$ if the
corresponding massless fields $\bphi^a$ and $\sigma^a$ arise from the
$U(n_i)$ factor of the gauge group. Finally, $\bnabla\times{\bf
W}_{ab}=\bnabla g_{ab}$. When $n_i=1$ for all $i$, the metric is
complete and exact. For $n_i\geq 2$, the metric has singularities
which are resolved by instanton corrections. In this case, the
resulting metric is no longer of the toric form \eqn{coul} and, in
particular, the tri-holomorphic isometries arising from constant
shifts in $\sigma^a$ are broken by the instanton corrections.

For the special case $n_0=0$, the gauge group ${\cal G}$ is based on
the non-affine Dynkin diagram and the above metric coincides with the
asymptotic metric on the moduli space of monopoles of gauge group $H$,
with magnetic charge $\vec{g}=\sum_{i=1}^rn_i\vec{\beta}_i$
\cite{threemon}. The generalisation of this result is that for
$n_0\neq 0$, Eq.~\eqn{coul} agrees with the asymptotic metric on the
moduli space of calorons. For gauge theories built on $A_r$ Dynkin
diagrams, this fact was pointed out some time ago \cite{caloron}.  The
gauge coupling constants are related to the masses of the monopoles
which, in turn, are determined by the VEV of the Wilson line, together
with the radius, R, of the $S^1$. In particular, we
have
\be \frac{1}{2\pi
R}=\sum_{i=0}^r\frac{d_i}{e^2_i} \; .
\label{r}\ee
(Here the dimension of both sides  is matched by  a
reference scale needed for comparing two scales of the equation.)
One may ask whether there is any natural deformation of the Coulomb
branch metric \eqn{coul} within the context of the three dimensional
gauge theory.  Such deformations arise through mass terms for the
hypermultiplets and, as we shall see, lead to moduli spaces of
calorons on non-commutative spaces. This observation has been previously 
made in \cite{kapseth}. It is simple to see that for the
$D$ and $E$ Dynkin diagrams, all such mass terms may be absorbed
through shifts of the vector multiplet scalars $\bphi$. This appears
to tie in with the fact that there is no known non-commutative version
of $SO(2r)$ and $E_r$ gauge theories. Similarly, for $A_r$ Dynkin
diagrams, no such mass term is allowed if, say, $n_0=0$. Again, this
dove-tails nicely with the fact that the monopole moduli space remains
unchanged in the non-commutative theory.  However, if $n_i\neq 0$ for
all $i=0,\cdots,r$, then we may use the freedom to shift ${\bphi}$ to
ensure that all but one of the bi-fundamental hypermultiplets has zero
mass. This final hypermultiplet which, for $r\geq 2$, we may choose to
be the one transforming in the $({\bf n}_0,\bar{\bf n}_1)$
representation, has a triplet of mass parameters ${\bf m}$. The
Coulomb branch metric for $r\geq 2$ is then given by \eqn{coul}, with
\beqn
 g_{aa} &=& \frac{1}{\tilde{e}_a^2}-\sum_{b\neq
a}\frac{\vec{\alpha}_a\cdot\vec{\alpha}_b} {|\bphi^a-\bphi^b+{\bf
m}_{ab}|} \; , \nonumber\\
g_{ab}&=&\frac{\vec{\alpha}_a\cdot\vec{\alpha}_b}{|\bphi^a-\bphi^b+{\bf
m}_{ab}|}\ \ \ \ (a\neq b) \; , \nonumber
\eeqn
with 
\beqn {
\bf m}_{ab}&=&+{\bf m}\ \
\ {\rm if}\ \vec{\alpha}_a=\vec{\beta}_0\ {\rm and}\
\vec{\alpha}_b=\vec{\beta}_1 \; , \nonumber \\ 
{\bf m}_{ab}&=&-{\bf m}\ \ \ {\rm
if} \ \vec{\alpha}_a=\vec{\beta}_1\ {\rm and}\
\vec{\alpha}_b=\vec{\beta}_0 \; , \nonumber \\ 
{\bf m}_{ab}&=&\ 0\ \ \ \ \ \ {\rm otherwise} \; . \nonumber
\eeqn
This is the asymptotic metric on the moduli space of
non-commutative calorons.  Note that for the case of $r=1$ ($SU(2)$),
the metric does not take the above form: there are now two
hypermultiplets in the $({\bf n}_0,\bar{\bf n}_1)$ representation, of
which one is massless and one is of mass ${\bf m}$. The corresponding
metric is given by, 
\beqn
 g_{aa} &=& \frac{1}{\tilde{e}_a^2}-\sum_{b\neq
a}\ft12
{\vec{\alpha}_a\cdot\vec{\alpha}_b}\left(\frac{1}{|\bphi^a-\bphi^b+{\bf
m}_{ab}|} +\frac{1}{|\bphi^a-\bphi^b|}\right) \; ,  \nonumber \\
g_{ab}&=&\ft12{\vec{\alpha}_a\cdot\vec{\alpha}_b}\left(\frac{1}{|\bphi^a
-\bphi^b+{\bf m}_{ab}|} +\frac{1}{|\bphi^a-\bphi^b|}\right)\ \ \ \ 
(a\neq b) \;  \nonumber
 \eeqn
with ${\bf m}_{ab}$ given as above.

We now turn to the main focus of this paper: non-commutative calorons
in the $U(1)$ gauge theory. Note that, while the above construction
was for calorons in gauge group $U(r+1)$ for $r\geq 1$, it may be
extended to the $r=0$ case.  The gauge theory describing $N$ $U(1)$
calorons is a $U(N)$ gauge group with a single adjoint hypermultiplet
of mass ${\bf m}$. The perturbative metric is this time given by, 
\beqn
g_{aa} &=&
\frac{1}{\tilde{e}^2}-\sum_{b>a}\left(\frac{2}{|\bphi^a-\bphi^b|} -
\frac{1}{|\bphi^a-\bphi^b+{\bf m}|}-\frac{1}{|\bphi^a-\bphi^b-{\bf
m}|} \right)\; , \nonumber \\
g_{ab}&=&\left(\frac{1}{|\bphi^a-\bphi^b|}-\frac{1}{|\bphi^a-\bphi^b+{\bf
m}|}\right) \ \ \ \ (a\neq b)  \; . 
\label{uncal}  \eeqn
The metric has a single $U(1)$
isometry which rotates $\bphi$ fixing ${\bf m}$. The isometry is
holomorphic with respect to a preferred complex structure. Note that
the tri-holomorphic $U(1)$ isometry of the perturbative metric that
results from shifts of $\sigma$ is broken by instanton effects. For
$N=2$, one may factor out the center-of-mass to find the metric
\eqn{asympt} with the appropriate normalization of the coefficients
and the identification ${\bf m}\sim(0,0,\zeta/2R)$.

It is interesting to examine various limits of the above metric. For
instance, as ${\bf m}\rightarrow\infty$, with $e^2$ held fixed, the
$U(1)$ holomorphic isometry is enhanced to an $SU(2)$ isometry which
rotates the three complex structures. In this limit, the metric
becomes the moduli space of $N$ monopoles in $SU(2)$ gauge group
which, for $N=2$ is the Atiyah-Hitchin manifold.  This is the limit
where two plus heads of two dipoles are close to each other and
separated far from two negative heads of two dipoles, and so the
moduli space becomes that of two indentical monopoles.  In general, in
the limit of large non-commutivity, the moduli space of $N$
non-commutative $U(k)$ calorons can take several different form. In
the above metric (\ref{uncal}) in the infinite ${\bf m}$ limit becomes
the moduli space of $(N,N,\cdots,N)$ $SU(k+1)$ monopoles. This is the
limit where dipoles are arranged so that one group negative heads of
$U(1)$ charge is close to another group of positive heads of the same
charge. One could take another limit where all positive heads are
together. In this case the positive heads of different $U(1)$ would
not interact and so one ends of $k$ copies of the $N$ moduli space of
$SU(2)$ monopoles.

A less well understood limit is $e^2\rightarrow\infty$. In this
regime, the non-perturbative instanton corrections are no longer
sub-leading and we have little control over the metric. Nevertheless,
using equation \eqn{r}, we see that this is the decompactification
limit in which the Coulomb branch becomes the moduli space of $N$
$U(1)$ instantons. {}From the results of section 3, we therefore find
the result that the Coulomb branch of $SU(2)$ with a single massive
adjoint hypermultiplet becomes Eguchi-Hanson in the strong coupling
limit. It would be interesting re-derive this result using purely
field theory methods. In particular, in this limit the metric develops
an enhanced $SU(2)$ tri-holomorphic isometry.  Such symmetry
enhancement, while not at all uncommon \cite{is}, remains poorly
understood from field theoretic considerations.  It may be worth
noting however that usually the symmetry group is enhanced from
abelian factors to a non-abelian group of the same rank. In the
present case, there is no tri-holomorphic isometry of the Coulomb
branch metric for finite coupling constant.

Finally, let us discuss the caloron decompactification limit, 
$R\rightarrow \infty$, for general simply-laced gauge group $H$. 
As explained in \cite{kimyeong}, this leads to a finite action 
instanton solution of charge 
$k$ precisely if the magnetic charges are chosen such that $n_i=kd_i$. 
Using \eqn{r}, the decompactification limit of the caloron 
moduli space coincides with the strong coupling limit of 
the three dimensional gauge theory, $e_i\rightarrow\infty$. 
In fact, the strong coupling limit of gauge theories very 
similar to these have been arisen previously in the study 
of mirror symmetry \cite{is, berk}. In particular, the 
authors of \cite{is} discuss the $k=1$ case. However, 
we find a subtle disagreement with the $k\geq 2$ 
generalisation discussed in \cite{berk}, the resolution 
of which leads to a  intriguing new prediction for the 
strong coupling dynamics of the gauge theory. Let us firstly 
restrict attention once more to the $A_r$ gauge groups. We 
will make a few comments on the $D$ and $E$ cases at the end. 
The general mirror pairs discussed in \cite{berk} are 

{\bf Theory A:} $U(k)$ gauge group with a $N$ fundamental
hypermultiplets and a single adjoint hypermultiplet. There is a
Fayet-Iliopoulos parameter (FI) $\zeta$, an adjoint mass parameter $M$
and fundamental mass parameters $m_i$, $i=1,\cdots,N$, where we may
choose $\sum_{i=1}^Nm_i=0$. The gauge coupling constant is $e$.  \\
{\bf Theory B:} $\prod_{i=1}^N U(k_i)$ gauge group where $k_i=k$ for
all $i=1,\cdots,N$. There is a bi-fundamental hypermultiplet in each
$(\bar{k}_i,k_{i+1})$ representation for $i=1,\cdots,N$ \footnote{we
set $k_{N+1}=k_1$}. The gauge group $U(k_1)$ has a further fundamental
hypermultiplet transforming in $k_1$. There is a single mass parameter
$\tilde{m}$ for this additional hypermultiplet transforming in the
$(\bar{k}_N,k_1)$ representation, while all remaining hypermultiplets
have no mass parameters. There are also $N$ FI parameters,
$\tilde{\zeta}_i$.  The gauge coupling constants are $\tilde{e}_i$.

Note that Theory B is very similar to the theory that we introduced
whose Coulomb branch coincides with the caloron moduli space if one
chooses $n_i=k_i=k$. The two theories differ however by the presence
of the fundamental hypermultiplet in Theory B.  The Higgs branches of
Theory A and Theory B, which arise as hyperkahler quotients, are both
known in the mathematics literature.  They exist fully only when
$M=m_i=\tilde{m}=0$, so let us concentrate on the theories with this
restriction. Firstly, as discussed in section 2, the Higgs branch of
Theory A is the moduli space of $k$ instantons in a $U(N)$ gauge
theory, ${\cal M}_{k,U(N)}$.  Turning on the FI parameter $\zeta$
deforms this space to the moduli space of instantons in a
non-commutative gauge theory.

In contrast, the Higgs branch of Theory B is the Hilbert scheme 
of $k$ points in the $A_{N-1}$ ALE space. We denote this by 
$Hilb^k(A_{N-1})$. This space is the smooth resolution of the 
symmetric product $S^k(A_{N-1})=A_{N-1}^k/S_k$. The 
differences in the FI parameters, 
$\tilde{\zeta}_i-\tilde{\zeta}_{i+1}$, determine the size of the 
two-spheres which sit inside the blown-up $A_{N-1}$ ALE. The sum of the 
FI parameters, $\sum_{i}\tilde{\zeta}_i$ determines the blow-up 
of the quotient singularities. 

To summarise, 
\beq
{\rm Higgs}_A = {\cal M}_{k,U(N)}\; , \ \ \ \ \ \ \
{\rm Higgs}_B = Hilb^k(A_{N-1})\; .
\nn\eeq
The statement of mirror symmetry in these theories is that in the 
infra-red limit $e\rightarrow\infty$ and 
$\tilde{e}_i\rightarrow\infty$, 
the Coulomb branches are also given by these spaces,
\beqn
{\rm Coulomb}_A &=&  Hilb^k(A_{N-1})\; , \label{cbina}\\
{\rm Coulomb}_B &=& {\cal M}_{k,U(N)} \;, 
\label{cbin}\eeqn
where the mirror map between parameters is given by,
\beq
M=\sum_{i=1}^N\tilde{\zeta}_i\;  , \ \ \ \zeta=\tilde{m}\; ,\ \ \ 
m_i=\tilde{\zeta}_i-M/N \; .
\nn\eeq
Note in particular for $N=1$, the theory is self-mirror and 
the above statement reduces to ${\cal M}_{k,U(1)}=Hilb^k({\bf R}^4)$

While much evidence was presented in \cite{berk} for the first of 
these claims \eqn{cbina}, much less is known about the latter \eqn{cbin}. 
In particular, explicit comparisons of the metrics are 
hard to perform as instanton corrections are no longer 
sub-leading in the strong coupling limit. Nevertheless, 
if we accept the mirror symmetry proposal of \cite{berk}, 
we are  left with a situation in which both Theory 
B and the caloron theory introduced at the beginning 
of this chapter both yield the instanton moduli space 
in the strong coupling limit. While it is obvious 
that at finite $e_i$, the two Coulomb branches differ, 
it appears that the presence of the additional fundamental 
hypermultiplet of Theory B plays no role in the strong coupling 
dynamics. While such statements are known to be 
true for abelian theories (see for example \cite{kaps}), 
the correspondence here requires remarkable cancellations 
between the instanton expansion of the two theories. 

In fact, evidence for this correspondence arises from the 
brane construction of these models \cite{berk2}. One can consider 
Theory A as the world-volume dynamics of $k$ D3-branes in directions 
0126 with the $x^6$ direction compactified. There is a single 
NS5-brane in the 012345 direction and $N$ D5-branes in the 
012789 directions. As usual, performing an S-duality results in 
Theory B. However, one could just as easily construct Theory 
A without the NS5-brane. S-duality then yields the caloron theory 
introduced at the beginning of this section.  
While the field content is unchanged, the removal of the 
NS5-brane does remove certain parameters. In Theory A, 
the adjoint mass  parameter is not available once the NS5-brane 
is removed. {}From the mirror map, the sum of FI paramters of 
Theory B is unavailable in the dual picture.

While the above discussion was for the $A_r$ group of 
theories, a similar discussion holds for the $D_r$ 
group. For the $E_r$ series, the mirror theory (i.e 
the equivalent of Theory A) has no known Lagrangian 
description and the point is moot. 

Finally we comment that in abelian theories, the possibility of adding 
a hypermultiplet without changing the strong coupling 
dynamics has been used in \cite{kaps} to give an elegant formalism 
of mirror symmetry in these theories. It would be intriguing if the 
observation here would yield a similar construction for non-abelian 
theories.

\subsection*{Acknowledgements}

We would like to thank M. Aganagic, A. Hanany, K. Hori, A. Karch and
P. Yi for useful discussions. K.L. is supported in part by KOSEF 1998
Interdisciplinary Research Grant 98-07-02-07-01-5.  D.T. would like to
thank KIAS for hospitality while this work was initiated, and CTP, MIT
for hospitality while it was completed. D.T. is supported by an EPSRC
fellowship. S.Y. is supported in part by the BK21 project of Education
Ministry.

\end{document}